 \documentclass[preprint]{aastex62}

\received{August 25, 2019}
\accepted{September 30, 2019}

\submitjournal{ApJ}

\shorttitle{Chromospheric UV bursts}

\begin{document}

\title{Chromospheric UV bursts and turbulent driven magnetic reconnection}

\correspondingauthor{Pin Wu}
\email{nanopenny@gmail.com}

\author[0000-0002-3366-8096]{Pin Wu}
\affil{School of Mathematics and Physics \\
Queen's University Belfast \\
Belfast, BT7 1NN, The United Kingdom}
\affil{Max-Planck-Institut fuer Kernphysik\\
Saupfercheckweg 1\\
Heidelberg, D-69117, Germany}

\begin{abstract}
We use Interface Region Imaging Spectrograph (IRIS) spacecraft data to study a group of Chromospheric ultraviolet bursts (UVBs) associated with an active region. We classify the UVBs into two types: smaller ones that can only be measured once by the scanning slit, and larger UVBs that are measured twice by the slit. The UVBs' optically thin Si IV 1402.77 \r{A} line profiles are studied intensively. By fitting the smaller UVBs' lines with 1-2 Gaussians, we obtain a variety of line-of-sight flow measurements that hint various 3-D orientations of small scale magnetic reconnections, each associated with a UVB. The larger UVBs are, however, unique in a way that they each have two sets of measurements at two slit locations. This makes it possible to unambiguously detect two oppositely directed heated flows jetting out of a single UVB, a signature of magnetic reconnection operating at the heart of the UVB. Here we report on the first of such an observation. Additionally, all the optically thin Si IV 1402.77 \r{A} line profiles from those UVBs consistently demonstrate excessive broadening, an order of magnitude larger than would be expected from thermal broadening, suggesting that those small scale reconnections could be driven by large scale (macro-scale) turbulence in the active region.

\end{abstract}

\keywords{Chromosphere --- Transition Region ---
UV burst --- reconnection --- turbulence}


\section{Introduction} \label{sec:intro}

\noindent Solar Ultraviolet bursts (UVBs) are intense, short-lived, and compact brightening events at $\sim 10^5$ K first observed \cite[]{Peter2014} by the Interface Region Imaging Spectrograph (IRIS) launched in 2013 \cite[]{DePontieu2014}. Providing a link between photospheric and chromospheric magnetic activities and the heating of the corona, UVBs have since been an active research subject in the field of the lower solar atmosphere remote sensing and have recently been reviewed by \cite{Young2018}. Although it is difficult to identify magnetic reconnection directly through remote sensing, several papers have attributed UV bursts to small-scale magnetic reconnection events occurring at photospheric and/or chromospheric heights \cite[e.g.,][]{Nobrega-Siverio2017, Voort2017, Young2018, Hansteen2017} as well as plasmoid formations \cite[e.g.,][]{Innes2015} through careful examinations and simulations. 

UVBs are of active regions origins \citep{Young2018}. An active region on the Sun is a region where the Sun's magnetic field is significantly disturbed. Dark sunspots are visual indicators of active regions. The strong magnetic disturbance in active regions can drive large macro-scale turbulence, which evolves and decays as it is convected outward from the Sun with the solar wind flow. Statistical theories of turbulent relaxation predicts that the fluid will relax to spatially non-uniform states, which manifest itself as intermittent coherent structures where most of the dynamic energy dissipates into heat. In a kinetic plasmas, this process unavoidably leads to the formation of reconnecting current sheets, as seen in magnetohydrodynamics (MHD) \cite[e.g.,][]{Matthaeus1986, Servidio2012} and fully electromagnetic particle-in-cell (PIC) simulations \cite[e.g.,][]{Wu2013}\footnote{A simulation movie from the cited paper can be found on \url{https://youtu.be/DAtLhKrF37o}} of decaying turbulence initiated with large scale broadband random phase fluctuations. Turbulent driven reconnection was also observed in the solar wind \cite[e.g.][]{Retino2007}. On the other hand, micro-scale turbulence can develop within reconnection and mediate reconnection rate, as initially suggested by \cite{Lazarian1999}, seen in PIC simulations \cite[e.g.,][]{Daughton2011, Dahlin2017}, and observed by the Magnetospheric Multiscale (MMS) mission \citep{Voros2017}.

Observationally, UVBs share similar phenomena with photospheric Ellerman Bombs (EBs) \citep{Ellerman1917} of the $H_{\alpha}$ line in the optical range, which are often attributed to reconnection \citep{Georgoulis2002}. Correlation of EBs and UVBs are found \citep{Vissers2015, Kim2015, Tian2016} and it is likely that the underlying physics of EBs and UVBs are related. However, UVBs and EBs are not necessarily always co-spatial with one another \cite{Tian2016}, probably because of the solar rotation and the solar wind convection from EBs height ($\sim 1000$ km) in the photosphere to UVBs height in the chromosphere and the transition region.

Another observationally transient feature that is similar to UVBs is the mini flare. Although, flares are located in the corona and will have loop emission in the 94 \r{A} and 131 \r{A} channels \citep{Young2018} of the Solar Dynamics Observatory (SDO) mission's Atmospheric Imaging Assembly (AIA) instrument, unlike the UVBs.

In this paper, we study a group of an active region's UVBs (2017-09-03 15:44-15:50) in the chromosphere and the transition region. Spectral line profiles of the optically thin Si V 1402.77\r{A} lines show that the UVBs are associated with various flow velocities, revealing that reconnections operate in various directions in 3-D. We would like to highlight our first unambiguous detection, through two measurements of the same UVB at two slit positions, of two oppositely directed heated flows jetting out of a single UVB, a signature of magnetic reconnection operating at the heart of the UVB. 

The excessive non-thermal broadenings of the optically thin Si IV line profiles suggest that the UVB related small scale reconnections are driven by turbulence in the entire active region, where small scale reconnecting current sheets are spontaneously and intermittently forming and evolving in a non-linear manner. These fine scale reconnection current sheets are locally enhanced heating/dissipation sites \cite[]{wu2013b} as well as sources of local reconnection outflows.

\section{Methods, Results, and Discussions} \label{sec:obs}

\noindent 

The IRIS mission has one instrument, an ultraviolet (UV) imaging spectrometer. The spectrometer makes high spatial and temporal resolution observations of selected regions of the solar chromosphere, the layer of the solar atmosphere just above the photosphere at $\sim$ 4000 K, and the transition region (TR), the region between the chromosphere and the corona where the temperature rises drastically. The observations are made in two channels: the far-UV (FUV: from 1332 to 1407 \r{A}) and the near-UV (NUV: from 2783 to 2835 \r{A}). The spectrometer's spatial resolution is 0.33 arcsec ($\sim$ 240 km) and its wavelength resolution is 0.026 \r{A} in the FUV range studied in this paper. There are two sets of observations, the slit-jaw imager (SJI) images and the raster spectrogram (SG) datasets \cite[]{DePontieu2014}.

\begin{figure}
\plotone{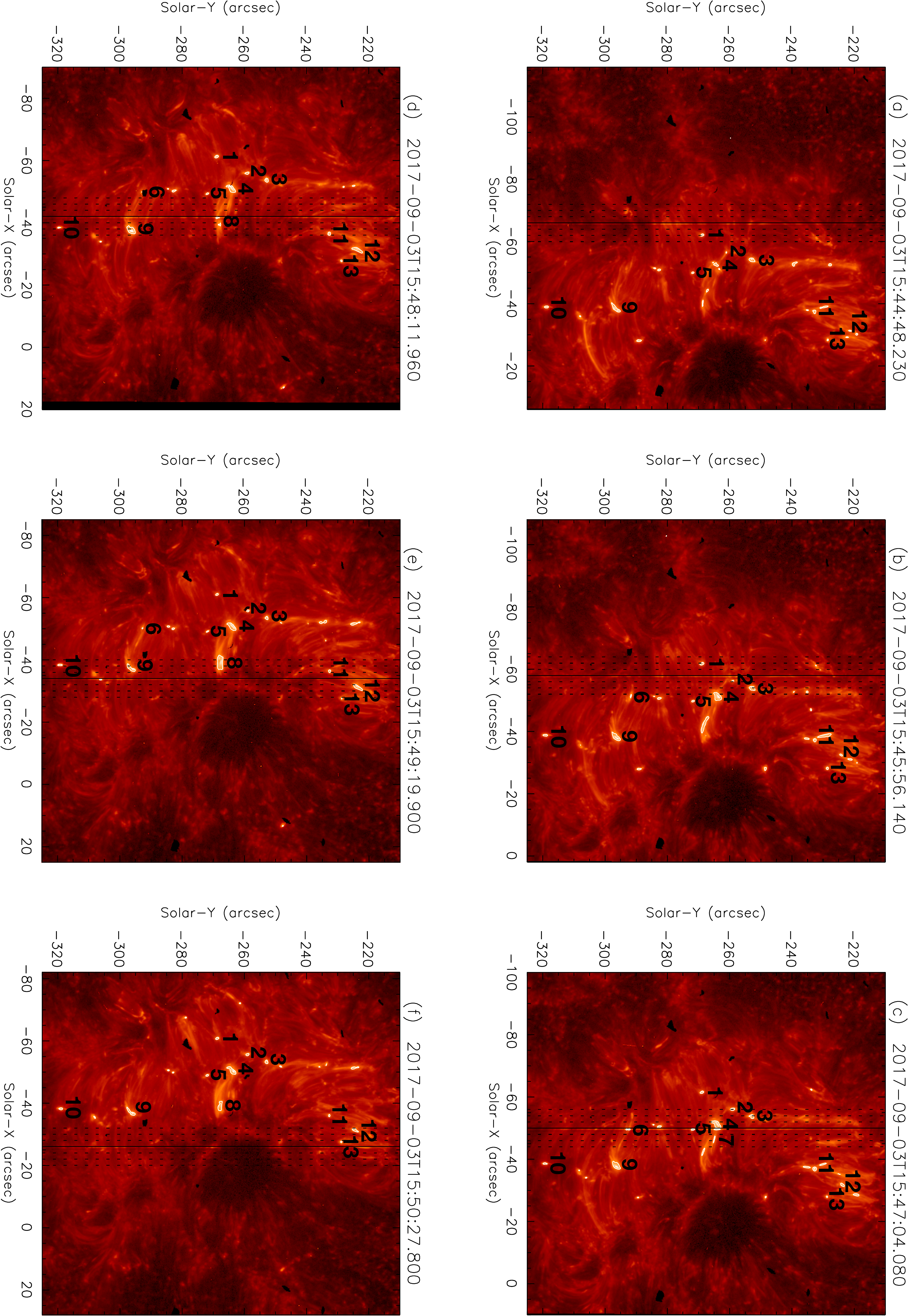}
\caption{Panel (a)-(f): time series of IRIS SJI 1400 \r{A} logarithmic intensity (DN/s). All panels are normalized between 0.1-1000 DN/s, with black being the lowest intensity and white being the highest intensity. The preceding/next 3 slit positions for each plot are marked with dotted lines.  The contours mark the identified UVBs where the intensity is a factor of 24 above the image median. We enumerate the UVBs that have been measured by the slit. Each UVB's number is mark on top of itself. \label{fig:fig1}}
\end{figure}

We identify UVBs from an active region using the IRIS SJI 1400 \r{A} images and following the criteria outlined in \cite{Young2018}. The examined period is from 15:44:48.230 to 15:50:27.800 UT on 2017-09-03, as shown in Figure \ref{fig:fig1} Panels (a)-(f) time series. For this observation period, the SJI images were taken every 68 seconds however the slit made spectrogram (SG) measurements as it scanned through the observed region every 68/4=17 seconds. Therefore, for each SJI image at an observed time, the slit position is plotted in black solid line and the preceding/next three slit positions are plotted with dotted lines. We subsequently define a slit position notation used throughout this paper for convenience. Let ``a+/-1(2,3)" denotes the slit position marked by the 1st (2nd, 3rd) dotted line to the right/left of the slit position in Figure \ref{fig:fig1} Panel (a). We identify UVBs using contours that encircle areas with intensities at a factor of 24 above the image median, as classified by a recent review \cite[]{Young2018}. We are able to analyse a UVB's spectral line profiles if the slit ever made a measurement upon passing the UVB. In Table \ref{tab:tab1}, we enumerate such measurable UVBs and their slit positions. Due to their small sizes, most UVBs were only measured once by the scanning slit. We will call those UVBs smaller UVBs throughout the text. On the contrary, the slightly larger UVBs 4, 8, and 12, were measured twice by the scanning slit, first at their left edges and then at their right edges 17 seconds afterwards. We will call those UVBs larger UVBs (larger than $\sim$2 arcsec, or $\sim$1442 km) throughout the text. A close examination of Figure \ref{fig:fig1} discloses that all UVBs evolve with time and most UVBs last for several minutes. In particular, we note that UVB 8, when viewed in Figure \ref{fig:fig1} Panel (e), appears to be one larger UVB, while when it is viewed in Figure \ref{fig:fig1} Panel (d) 68 seconds earlier, it appears to be two closely related but seemingly separate UVBs. UVB 7 appear to last for the shortest time: slightly more than one minute.

\begin{deluxetable*}{cccccc}[b!]
\tablecaption{UVBs with spectrogram (SG) measurements \label{tab:tab1}}
\tablecolumns{4}
\tablenum{1}
\tablewidth{0pt}
\tablehead{
\colhead{UVB number} &
\colhead{SG data time} &
\colhead{Slit position\tablenotemark{a}} & 
\colhead{Solar-Y\tablenotemark{b}} & 
\colhead{} \\
\colhead{} & \colhead{(hh:mm:ss)} & 
\colhead{} & \colhead{(arcsec)} & \colhead{}
}
\startdata
1 & 15:45:22.410 & a+2/b-2 & -269.3  \\
2 & 15:46:13.460 & b+1/c-3 &  -261.6\\
3 & 15:46:30.330 & b+2/c-2 &   -253.0\\
4 & 15:46:47.310  & b+3 &  -264.4\\
  &  15:47:04.160 &  c &  -264.8\\
5 & 15:47:04.160 & c &  -272.3& \\
6 & 15:47:04.160 & c &  -292.5& \\
7 & 15:47:38.210 & c+2 & -265.6\\
8 & 15:48:12.030 & e-3 & -268.1\\
 &  15:48:29.310 & e-2  & -268.3\\
9 & 15:48:46.180 & d+2/e-2  & -297.2 \\
10 & 15:48:46.180 & d+2/e-2 & -320.3 \\
11 & 15:49:19.970 & d+3/e-1 & -233.0 \\
12 & 15:49:37.210  & e+1  & -225.0\\
 &  15:49.54.050 & e+2 & -220.0\\
13 & 15:50:11.020 & e+3/f-1 & -229.2\\
\enddata
\tablenotetext{a}{We refer to Figure \ref{fig:fig1} SJI image Panel (a)-(f) to denote slit position in this table. If there is no corresponding SJI image at a SG data time, the preceding/next SJI image is referred to. The slit position is subsequently noted as the +/- (1-3) slit position of that SJI image. For example, ``a+2" means that the corresponding slit position is along the 2nd dotted line to the right of the slit position in Figure \ref{fig:fig1} Panel (a), and ``b-2'' means that the corresponding slit position is along the 2nd dotted line to the left of the slit in Figure \ref{fig:fig1} Panel (b). Note that ``a+2" and ``b-2" mark the same slit position,  likewise ``b+1" and ``c-3" also mark the same slit position.}
\tablenotetext{b}{Each UVB's position along Solar-Y is taken from the centre (highest intensity) of the UVB as measured by the slit.}
\tablecomments{If more than one UVBs are observed at a time on the slit, we enumerate them from the top to the bottom along the slit.}
\end{deluxetable*}

\begin{figure}[ht!]
\includegraphics[width=\textwidth]{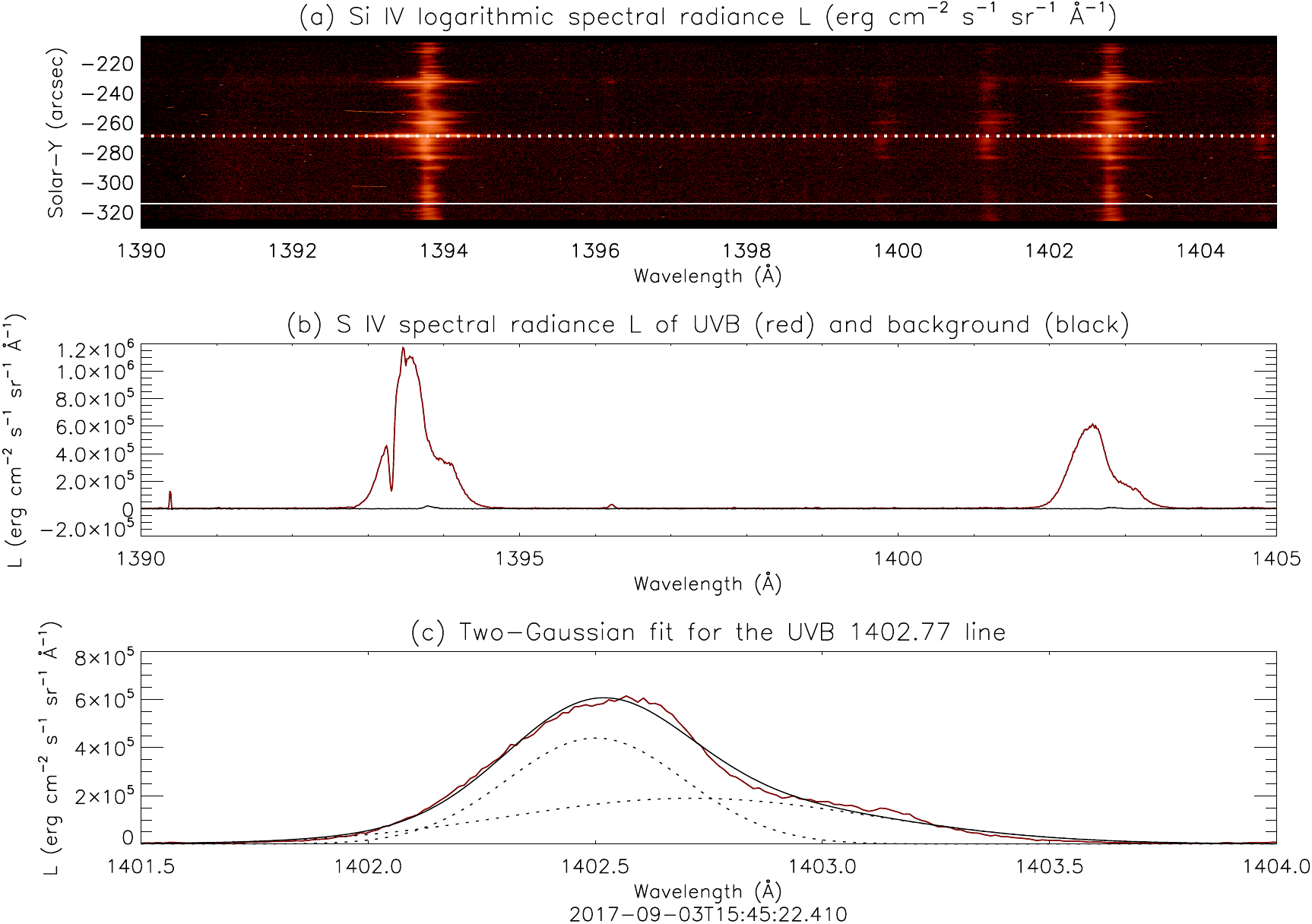}
\caption{Panel (a) is the Si IV spectral radiance of UVB 1. The dotted white line marks the UVB centre (Solar-Y=269.25742) as measured by the slit. The corresponding line profile is plotted in red Panel (b). The white line in Panel (a) marks the location (at Solar-Y=-313.50652 in this case) of a background measurement of the Si IV lines for comparison. This background emission is plotted in Panel (b) in black. Panel (c) Two-Gaussian fit for the UVB 1402.77 \r{A} line (red). The two Gaussians are plotted in dotted black lines. Their sum is plotted in the solid black line. The centres of the two Gaussians are marked by two vertical dotted black lines at 1402.50 and 1402.71 \r{A}, respectively. \label{fig:fig2}}
\end{figure}

Figure \ref{fig:fig2} is an overview of UVB 1's optically thin Si IV 1393.78/1402.77 \r{A} lines. Note that the Si IV 1394Å/1402Å ratio can be used to test whether the ion is formed under optically thin or thick conditions. A ratio close to 2, which is the case for the UVBs studied in this paper, is compatible with the lines being optically thin \cite[e.g.,][]{Peter2014, Yan2015}. However, optically thick lines can still have a ratio of about 2, so one cannot use the line ratio to conclusively state that the lines are optically thin. Panels (a) and (b) show that the UVB has remarkably intensive brightening and broadening of the Si IV lines, being orders of magnitude brighter in spectral radiance and significantly wider in wavelengths than the background Si IV emissions. In order to form the Si IV line, one needs a temperature of at least $\sim 80,000$ K \citep{Peter2014} or higher. The strong brightening of this line indicate strong heating of the UVB region plasmas to $\sim 10^5$K. Note that there is a significant absorption line (Ni II 1393.33 \r{A}) on the left wing of the 1393.78 \r{A} line, as seen before \cite[e.g.,][]{Young2018}. Sometimes small instrument effects may cause non-physical shift in lines, the Ni II 1393.33 \r{A} line is useful in a way that it provides a calibration reference point. We perform a two-Gaussian fit to the unblended 1402.77 \r{A} line which is not affected by the Ni II absorption. As seen in Panel (c), both Gaussians are blue-shifted, to 1402.50 and 1402.71 \r{A} respectively. These correspond to two jets of flows along the line-of-sight (LOS) at the bulk speeds of 58.12 km/s and 11.84 km/s respectively. The standard deviations of the two Gaussians are 0.19 and 0.40 \r{A}, respectively, which correspond to velocity broadening of $v_{BD}=\pm$41.61 km/s and $\pm$85.59 km/s respectively. Note that thermal broadening of Si IV at a temperature of $\sim 10^5$K is just about $ v_{th,BD}=\pm 4$ km/s (calculated from the thermal velocity along the line-of-sight $v_{th,LOS}=(2kT/m_{Si})^{1/2} \simeq 8 \rm km/s$). The excessive broadening of line profile observed here, so called non-thermal broadening $v_{nt,BD}=(v_{BD}^2-v_{th,BD}^2)^{1/2} \simeq v_{BD}$ (as $v_{BD} \gg v_{th, BD}$), is more than 10 times of pure thermal broadening. Even if the left-hand side component of Figure \ref{fig:fig2} (c) is made up of two components, the widest component will still be more than half of the width of it, which is still more than $\sim$ 5 times wider than pure thermal broadening. This is an indication of macro-scale turbulent motions, as \cite{Jeffrey2018} points out, opacity and pressure broadening are negligible in optically thin lines such as the Si IV lines here. 

Plots (not shown) from the other UVBs show the same trend of enhanced brightening and broadening of the Si IV lines as seen in Figure \ref{fig:fig2} for UVB 1. The spectral intensity enhancement in those profiles indicate that these regions all have temperatures $> 10^5 K$. We can fit all smaller UVBs' Si IV 1402.77 \r{A} lines with a multiple Gaussian fitting routine in a manner similar to the one described above. The Doppler shifts from these fits provide the line-of-sight velocities of the flows at these bursts. The results are listed in Table \ref{tab:tab2}. All smaller UVBs can be fitted with 1-2 broadened Gaussians. Particularly, each of UVB 3, 10, and 13 can be fitted with 2 Gaussians implying two oppositely directed flows along the line-of-sight. This is consistent with magnetic reconnection operating at those UVBs. 

\begin{figure}
\plotone{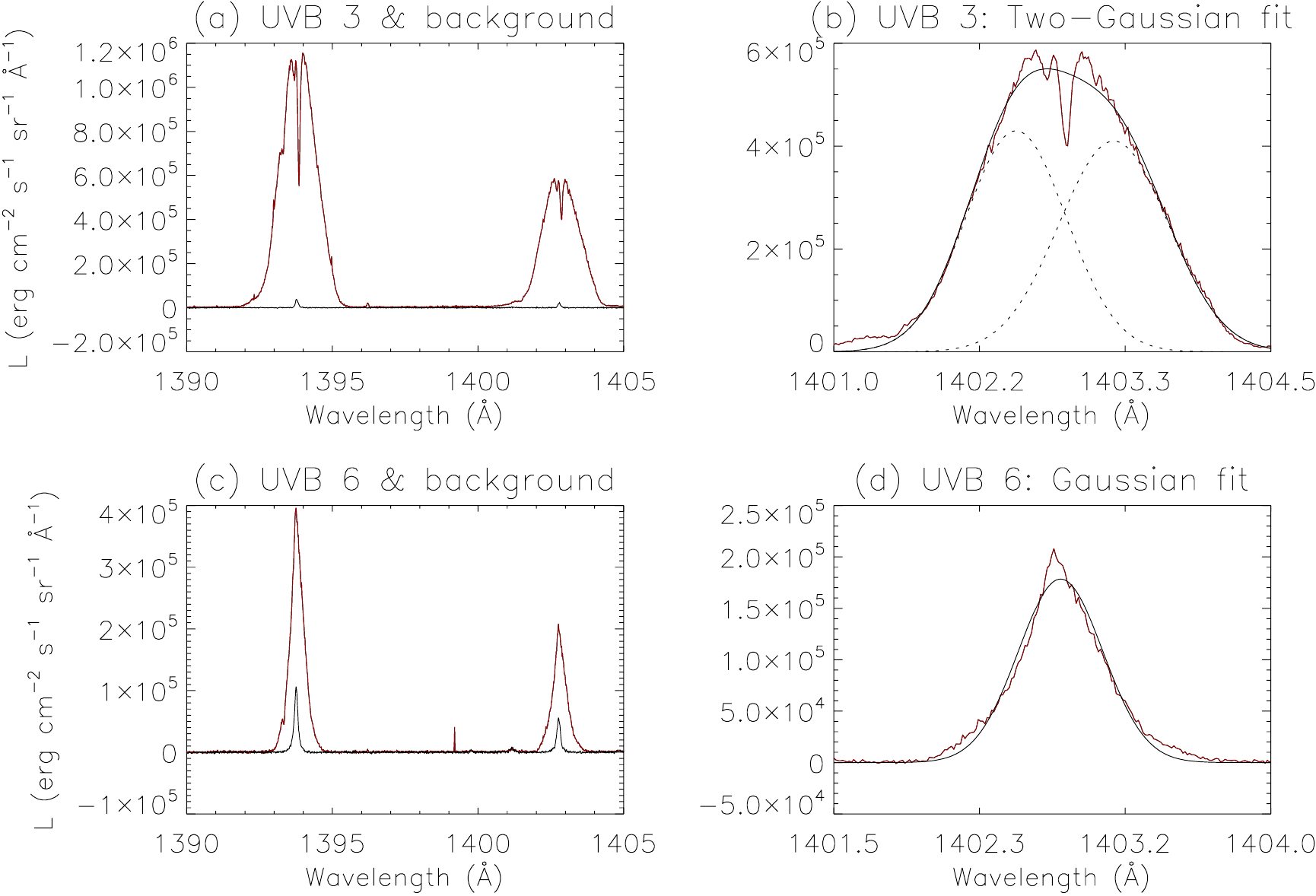}
\caption{Panel (a)/(c) are the Si IV line profiles of UVB 3/6 (red) and their backgrounds (black). Panel (b) is a two-Gaussian fit for the UVB 3's 1402.77 \r{A} line (red). The two Gaussians are plotted in dotted black lines. Their sum is plotted in the solid black line. Panel (d) is a Gaussian fit (black) for UVB 6 (red). The absorption in UVB 3, as also in UVB 13 (not shown). The emission in UVB 6 are apparent, as also in UVB 7 (not shown). \label{fig:fig3}}
\end{figure}

Peculiarly, two UVBs (3 \& 13) display apparent absorption at the un-shifted Si IV 1402.77 \r{A} line centres as also seen in the \cite{Hansteen2017} simulations, while two UVBs (6 \& 7) display apparent emission at the un-shifted line centres instead, as shown in Figure \ref{fig:fig3}. The absorptions are not wider than thermal broadening and are unlikely to affect the overall dynamics. They could be an indication of some background cold chromosphere plasmas stacked on top of the hot UVB plasmas, which present an equivalent optically thick effect. On the contrary, the emission in UVBs 6 and 7 may be due to reduced opacity, which is a result of increased temperature (heating) according to Kramer's opacity law \citep{Carroll1996}. Interesting enough, for UVBs 2, 6 (as seen in Figure \ref{fig:fig3}-d), 7, and 9, there are non-Gaussian line shape in the wings and the centre is more peaked. This non-Gaussianity could be due to additional moving components, or more likely, due to micro-scale turbulence generated by reconnection as in flare observations \cite[e.g.,][]{Jeffrey2018}. We observe that all UVBs display excessive non-thermal broadening, an order of magnitude broader than would be expected from mere thermal broadening, as a result of Doppler-shifted emissions from the macro-scale turbulent flows. It is thus compelling to remark that macro-scale turbulence can drive reconnections which in turn can generate micro-scale turbulence within the reconnection regions.

\begin{deluxetable*}{cccccccc}[b!]
\tablecaption{Multiple-Gaussian fit for the Si IV 1402.77 \r{A} line of smaller UVBs \label{tab:tab2}}
\tablecolumns{6}
\tablenum{2}
\tablewidth{0pt}
\tablehead{
\colhead{UVB number} &
\colhead{Number of Gaussians} &
\colhead{LOS velocity (km/s) \tablenotemark{a} } & 
\colhead{Velocity Broadening (km/s)} &
\colhead{Annotation\tablenotemark{b}} &
}
\startdata
1  & 2 & 58.1 &$\pm$46.6&\\
   & & 11.8 &$\pm$85.6&\\
2 & 1 & -8.1 &$\pm$54.6& \\
3 &  2 & 76.0 &$\pm$81.6& Absorption &\\
    &  & -88.3  &$\pm$100.0& \\
5  & 2 & 72.6 &$\pm$19.9 & \\
    &  & 1.7 &$\pm$53.6&\\
6 &  1 &  -5.8 &$\pm$51.8& Emission &\\
7 &  1 & 4.1 &$\pm$72.9& Emission &\\
9  & 1& 29.8 &$\pm$82.9& \\
10  & 2 & 24.4 &$\pm$65.9&\\
    &  &  -8.2 &$\pm$11.3&\\
11  & 2 & 15.0 &$\pm$53.0&\\
     &   & 11.5 &$\pm$14.0&\\
13 &  2 & 60.6 &$\pm$33.5& Absorption &\\
      &   & -33.5 & $\pm$37.8&\\
\enddata
\tablenotemark{a}{We use positive speed to denote flows with blue shifted spectra and negative speed to denoted flows with red shifted spectra.}
\tablenotemark{b}{We mark the UVBs that have obvious absorption or emission at the un-shifted Si IV 1402.77 \r{A} line centre.}
\end{deluxetable*}

\begin{figure}
\plotone{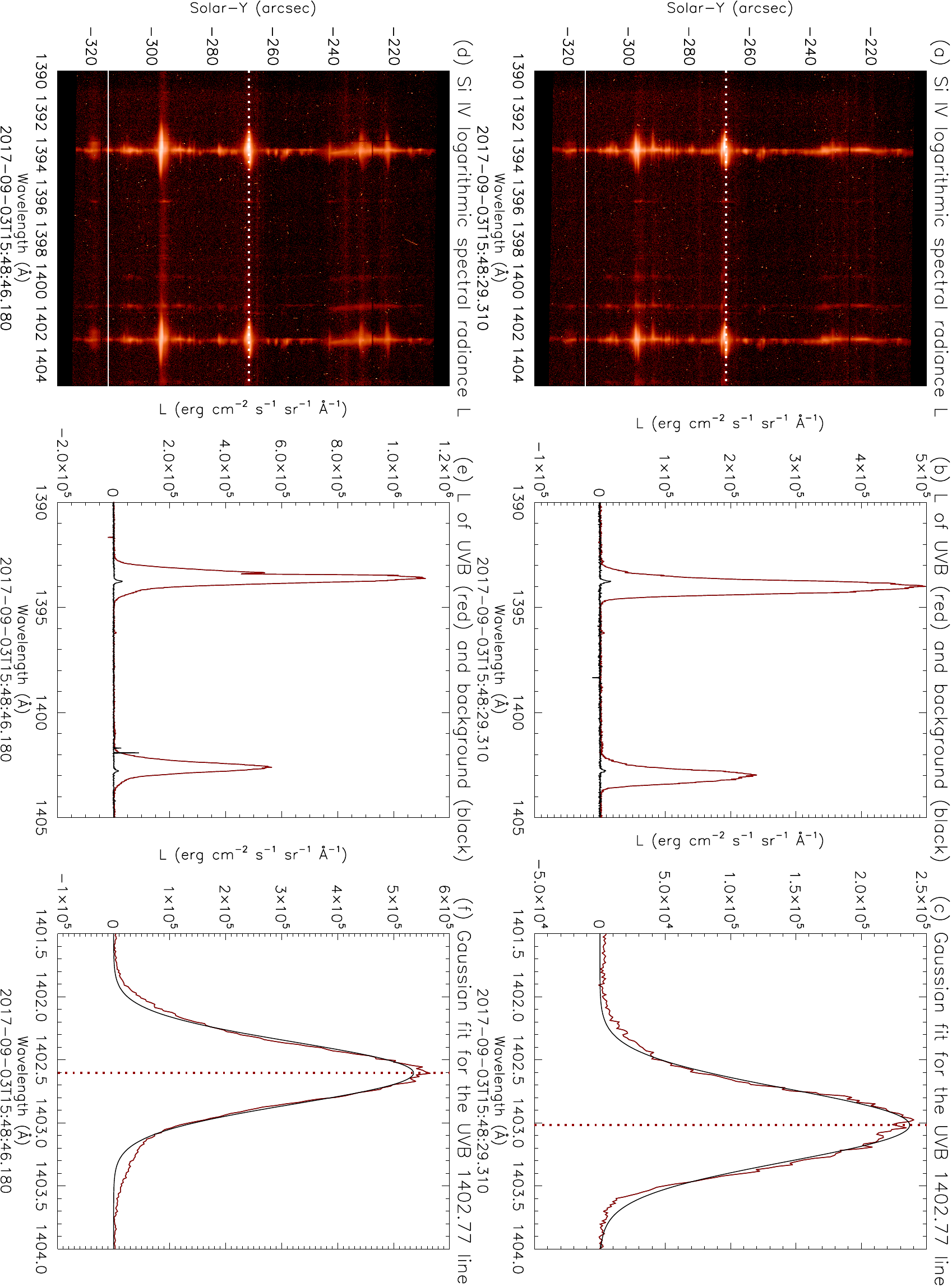}
\caption{UVB 4 Si IV lines. Panels (a), (b), and (c) are measurements made at slit position e-3 at 15:48:29.310 UT. Panels (d), (e), and (f) are measurements made at slit position e-2 at 15:48:46.180 UT. Panels (a) and (d) are Si IV logarithmic spectral radiance L ($erg \cdot cm^{-2} s^{-1} sr^{-1} \r{A}^{-1}$), where dotted white lines mark the centres of UVB 4 and the white lines mark the backgrounds. Panels (b) and (e) are the UVB centres line profiles (red) v.s. the background line profiles (black). Panel (c) and (e) are Gaussian fits (black) for the UVB's 1402.77 \r{A} lines (red). The centres of the two Gaussians are marked by two vertical dotted red lines at 1403.2 and 1402.6 \r{A}, respectively. \label{fig:fig4}}
\end{figure}

Similarly, we study the Si IV 1402.77 \r{A} lines of the larger UVBs 4, 8, and 12 using our multiple Gaussian fitting routine. The results are listed in Table \ref{tab:tab3}. Note that here we can measure these UVBs at two slit positions, mostly near the two edges of the UVBs. Again, all large UVBs demonstrate excessive broadening of line profiles as a result of macro-scale turbulent fluid motions that can drive reconnections. UVBs 4 and 8 also demonstrate non-Gaussian wings that imply micro-scale turbulence generated by reconnection. We find that most of these line profiles can be fitted with one Gaussian. In particular, we would like to single out UVB 8 for more detailed studies. Slightly to the left of the centre of Figure \ref{fig:fig1} Panel (e) and spanning slit positions e-3 and e-2, UVB 8 has an elongated shape. It stands out because its centres at both slit positions are at about the same Solar-Y location. A comparative study of the two slit positions' measurements is shown in Figure \ref{fig:fig4}. The measurement at slit position e-3 is made at 15:48:29.310 UT and plotted in Panels (a), (b), and (c). The measurement at slit position e-2 is made at 15:48:46.180 UT and plotted in Panels (d), (e), and (f). The most striking difference between these two measurements is that one measures a red shifted line as shown in Panel (c) and the other a blue shifted line as shown in Panel (f), corresponding to velocities of -53.0 km/s and 35.3 km/s along the line-of-sight, respectively (ref. Table \ref{tab:tab3}). The two oppositely directed heated flow measured at the two sides of the UVB are remarkably consistent with the physical picture of magnetic reconnection operating at the heart of this UVB. It is probably also worthwhile to mention that in the laboratory, the Swarthmore Spheromak Experiment (SSX) also measured bi-directional reconnection outflows using UV spectra (the C III 2297 \r{A} line) \cite[]{Brown2008}.

\begin{deluxetable*}{ccccccc}[b!]
\tablecaption{Multiple-Gaussian fit for the Si IV 1402.77 \r{A} line of the larger UVBs \label{tab:tab3}}
\tablecolumns{6}
\tablenum{3}
\tablewidth{0pt}
\tablehead{
\colhead{UVB number} &
\colhead{Slit position} &
\colhead{Number of Gaussians} &
\colhead{LOS velocity (km/s) } & 
\colhead{Velocity Broadening (km/s)} &
\colhead{Annotation} 
}
\startdata
4 & b+3& 2  & 71.6   & $\pm$89.5&\\
   &       &   & -98.2 &$\pm$39.0 &\\
   & c    &  1 & 21.1  & $\pm$40.1 & Emission &\\
8 & e-3 &  1 & -53.0 & $\pm$61.5&\\
   & e-2 & 1 & 35.3   & $\pm$52.1&\\
12&e+1 & 1 & -5.8    & $\pm$110.7 &\\
    &e+2 & 1 & 28.5    & $\pm$59.6&\\
\enddata
\tablecomments{This table has similar format as Table \ref{tab:tab2}. The only difference is that here each UVB has two sets of measurements, each at a separate slit position. Therefore, we include the slit position column to indicate the different observation locations.}
\end{deluxetable*}

We note that UVBs are also observed in Mg II and C II in the chromosphere. We chose to analyse the Si IV lines in this paper because the other lines are optically thick and hard to interpret. They may be a topic of future investigations. We also note that the sampled UVBs, smaller or larger, display a variety of flow speeds (as in Table \ref{tab:tab2} and \ref{tab:tab3}) along the line-of-sight, indicating that small-scale magnetic reconnection is occurring in all directions. This is consistent with a scenario in which those reconnections are driven by large macro-scale turbulence \cite[e.g.,][]{Matthaeus1986, Servidio2012, Wu2013} spanning this active region. This macro-scale turbulence manifest themselves as excessive non-thermal broadening an order of magnitude larger than would be expected from mere thermal broadening. Further, 19 out of 23 (equivalent of 83\%) of the turbulent broadenings indicate maximum LOS speeds (see Table \ref{tab:tab2} and \ref{tab:tab3} velocity broadening) larger than their respective LOS reconnection outflow speeds. Therefore, we suggest that the UVB-related reconnections here are results of a macro-scale turbulence driver. Finally, we would like to remark that the underpinning physics of Ellerman Bombs in the photosphere, UVBs in the chromosphere/transition region, and the mini flares in the corona could be similar. The validity/invalidity of this suggestion awaits more future investigations.


\section{Summary} \label{sec:sum}

The presence of ultraviolet bursts (UVBs) in the chromosphere and transition region indicate that heating in the solar atmosphere can be sporadically intense and spatially non-uniform.  Here we analyse a group of UVBs and their optically thin Si IV 1402.77 \r{A} line profiles. The 13 UVBs we examined are within one active region. For the first time, we are able to measure a few larger UVBs twice at two slit positions, from which we can directly and unambiguously measure two oppositely directed heated flows out of a UVB, a strong evidence of magnetic reconnection operating underneath those bursts. 

The UVB related flows are at temperatures of $\sim 10^{5}$K, orders of magnitude higher than average chromosphere temperature. Their line-of-sight (LOS) velocities are in the range of -98 to 76 km/s. The variation of the line-of-sight flow velocities in this range indicates that the flows are multi-directional and that the underlying reconnection processes are also multi-directional in 3-D. Such a group of multiple reconnections could be driven by large scale turbulence of the size of the active region in a manner described as turbulent driven reconnection \cite[e.g.,][]{Matthaeus1986, Servidio2012, Wu2013}. The strong magnetic disturbance of the active region provide a nature source of energy for generating this large scale turbulence. Another strong evidence of macro-scale turbulence is the excessive broadening (ranging from $\pm 11.3$ to $\pm 110.7$ km/s) of the Si IV 1402.77 \r{A} line profiles in all UVBs, about an order of magnitude larger than would be expected from mere thermal broadening. Moreover, most (83\%) of the inferred LOS turbulent flow speeds are larger than the related LOS reconnection outflow speeds, indicating that the turbulent driver is on scales larger than individual reconnections. Additionally, non-Gaussian wings of line profiles are seen in 6 out of the 13 UVBs we have studied , implying the generation of micro-scale turbulence by reconnections. We suggest that a portion of coronal heating are related to the consequential accumulation of heat due to macro-scale turbulent driven small-scale magnetic reconnection that also create the observed UVBs. Such a viewpoint is in accord with the \cite{Parker1988} vision of corona heating.

\acknowledgments

{IRIS is a NASA small explorer mission developed and operated by LMSAL with mission operations executed at NASA Ames Research center and major contributions to downlink communications funded by ESA and the Norwegian Space Centre. The IRIS data are publicly available from the Lockheed Martin Solar and Astrophysics Laboratory (LMSAL) website (\url{http://iris.lmsal.com/}). The open source SolarSoft code package (\url{http://www.lmsal.com/solarsoft/}) is used for the initial data processing. PW acknowledges the Science and Technology Facilities Council (STFC) and the Daphne Jackson Trust. PW thanks Professor Mihalis Mathioudakis and the anonymous referee for comments; Professor David Riley and Professor Marco Borghesi for their support; and Dr Arron Reid, Dr. Peter Young, Dr. Marc DeRosa, and Dr. Xudong Sun for discussions on data acquisition and processing. PW also wishes to express thanks to the Kavli Institute for Theoretical Physics (KITP) for hosting her visit to the ``Multiscale Phenomena in Plasma Astrophysics" program as well as funding her family's travel and child care, during which time she submitted and revised this paper. 




\bibliography{WuRef}




\end{document}